\documentstyle[referee]{l-aa} 
\input psfig.tex

.tfm    
\def\lsim{\lower.5ex\hbox{$\; \buildrel < \over \sim \;$}}
\def\gsim{\lower.5ex\hbox{$\; \buildrel > \over \sim \;$}}
 
\begin{document}
\title {Spectral Signature of Wind Generation From The Post-Shock Region in GRS1915+105 Accretion Disk}
\author{Sandip K. Chakrabarti$^{1,3}$, Sivakumar G. Manickam$^1$, Anuj K. Nandi$^1$  and A. R. Rao$^2$}
\institute {$^1$ S. N. Bose National Centre for Basic Sciences, Salt Lake, Calcutta 700098, India\\
$^2$Tata Institute of Fundamental Research, Homi Bhabha Road, Mumbai(Bombay) 400005, India\\
$^3$Centre for Space Physics, 114/v/1A Raja S.C. Mullick Rd., Calcutta 700047 India}
\offprints { S. K. Chakrabarti {\it chakraba@boson.bose.res.in}}
\date{Received ; accepted , }
\maketitle
\markboth{  }{}

\maketitle

\begin{abstract}
Accretion and outflows are common in systems which include black holes.
Especially important is the case of the well known micro-quasar 
GRS1915+105 in our own galaxy, where super-luminal outflows are 
detected. We present a few observation which are suggestive of an
outflow which is generated very close to the black hole, within a few tens 
of Schwarzschild radii. In the presence of mass loss (e.g., an
outflowing wind), the electron density of matter  within the centrifugal 
pressure supported region (which generates hard X-rays) goes down
and it is easier to cool these electrons by soft photons coming
from the Keplerian disk. If, on the other hand, the post-shock region gains 
mass from outside, the spectra would be harder. These properties of 
spectral `softening' of the low state and `hardening' of the high state
have been detected in several days of RXTE data of GRS1915+105 which we present here.

\keywords {X-rays: stars -- outflows -- winds -- 
Comptonization -- black hole physics -- GRS1915+105}
\end{abstract}

\noindent Submitted to Astronomy \& Astrophysics May, 2000; Second revision December 2000.

\medskip

\section{Introduction}

GRS1915+105 is a well known example of a micro-quasar which resembles 
a quasar in all aspects including having jets which are ejected at an 
apparent superluminal speed (Mirabel \& Rodriguez, 1994). Several 
observations have been made (Pooley et al. 1997 Eickenbery et al. 1998; 
Feroci et al., 1999; Fender et al. 2000a; Dhawan et al. 2000) which 
suggest that there is a distinct correlation between the 
Comptonizing region and the outflowing jets. For instance, 
Pooley et al (1997) showed that when the source goes from 
the burst-off state to the burst-on state via the X-ray dip, the radio 
oscillation starts. Eikenbery et al. (1998) pointed out that X-ray and IR flares are
triggered by the same events very close to the black hole. Using data of the
light curve (of the so-called $\beta$ class, Belloni et al. 2000a), they noted
that IR flares are associated with a spike formation and right after the 
spike matter may have come out as baby jets. Feroci et al. (1999)
associated disapearance of the inner part of the disk with the 
creation of radio flares. Fender et al. (2000a) find that only is hard states
there are continuous outflows while in very soft states the outflow could be missing.
Dhawan et al. (2000) observd that the starting 
time soft X-ray flares and radio flares are correlated
while hard X-ray is anti-correlated with radio intensity. Not only GRS 1915+105, similar
conclusions were drawn using observational data from using other black hole
candidates such as Cyg X-1 (Fender et al. 2000a), GX 339-4 (Fender et al. 1999a).

In this paper, while supporting the general observations made so far, 
we present spectral signatures which indicate activities associated with winds as well. The effect 
we discuss was predicted in 1998 (Chakrabarti, 1998a) and we find that 
in several days of spectra of GRS 1915+105 this effect is observed. While our search is not exhaustive,
we believe that this observation should be true in general and in other objects as well.
These would be reported elsewhere. Based on the theoretical paradigm and the observations
we believe that winds emerge from the centrifugal pressure supported innermost region of the 
disk, where hard X-rays are also emitted. To establish this we need to describe briefly the
black hole accretion/outflow paradigm as we know it today.

As matter enters into the horizon of a black hole, the accretion flow has 
to be necessarily supersonic and thus sub-Keplerian (Chakrabarti, 1990; 
hereafter C90, Chakrabarti \& Titarchuk, 1995, hereafter CT95; Chakrabarti 1996a), 
i.e., it must deviate from a Keplerian disk. The centrifugal barrier dominated boundary
layer (CENBOL for short) of a black hole (CT95) 
puffs up and intercepts soft photons from the pre-CENBOL
Keplerian disk. Depending on whether the Keplerian disk rate or the 
sub-Keplerian halo accretion rate dominates, the emerging spectra may be 
harder or softer (CT95, Chakrabarti 1997). It is seen that if the Keplerian disk rate is low (${\dot m}
=\frac{\dot M}{{\dot M}_{Edd}} \sim 0.01-0.3$) and the halo rate is high 
(${\dot m}\sim 1$) the spectra would be hard. In this case, the soft photons 
are fewer in number and they are unable to cool down the electrons in CENBOL by 
inverse Comptonization. With this disk composition, the black hole would be
in the so-called `hard state' (Tanaka \& Lewin, 1995; CT95).
On the other hand, if the Keplerian disk rate is high 
(${\dot m}  \sim 0.3-1$ or more), the spectrum would be soft. In this case, 
soft photons are profuse in number and since optical depth inside 
CENBOL crosses unity ($\tau >1$), the resulting spectrum is softened. 
The black hole would be said to be in a soft-state in this case. In fact, Chakrabarti (1997)
proved that even when the Keplerian rate is low, the spectrum could be soft
depending on the size and composition (whether there are sub-Keplerian flow or not)
of the Comptonizing region since they determine the degree of interception of soft photons
and the optical depth.

It is a general property of the advective disks that outflows must be present (C90)
since they are generalizations of the spherical Bondi (1952) flow which also showed
outgoing winds which we term as Parker-winds (Parker, 1982). It was therefore 
no surprise that when the anti-correlation of hard X-rays and radio intensity 
were first reported (Mirabel \& Rodrigues, 1994) immediate explanation was
that `` .. the collapse of fields in the funnel would cause destruction
of the inner part of the disk and formation of blobby
radio jets. Detailed observation of GRS 1915+105 shows these features. Since inner part of 
the accretion disk could literally disappear by this magnetic process, radio flares should
accompany reduction of X-ray flux in this objects." (1996b; see also Chakrabarti 1994).
It was subsequently shown, that the CENBOL, which is only around few tens of the
Schwarzschild radii large, could actually be responsible to form outflows and jets.
Chakrabarti (1998b, 1999a) estimated the ratio $R_{\dot m}$ of outflow rate to inflow 
rate and found that in the limit when the flow is isothermal till the sonic 
point of the outflow, $R_{\dot m}$ depends only on the compression ratio $R$ 
of the gas at the shock. Chakrabarti (1999b) showed that the presence 
or absence of outflows are directly related to the spectral states. In a truly soft state,
the shock is absent ($R\sim 1$) and no significant wind would be produced. 
In a hard state, shock could be strong, ($R\sim 5-7$). The outflow rate 
would be reduced but QPO may exist due to shock oscillation. 
For an intermediate shock (say $R \sim 2.5-3$), the outflow rate 
is large and the sonic surface in the outflow is close enough to be filled in 
quickly to create the burst-on/burst-off phenomnon and to
form blobby jets.
Subsequently, Das and Chakrabarti (1999) computed $R_{\dot m}$ rate 
more accurately using globally complete inflow and outflow transonic solutions 
and found that in some region of the parameter space the entire sub-Keplerian
region could be completely evacuated. 

While there is much confusion in the literature about the nomenclature of 
states, we  would like to define the states in a manner that is generally accepted
in the literature. (Some works where these definitions are given are: Tanaka \& Lewin, 1995;
CT95; Pooley et al. 1997; Honam et al., 2000; Belloni et al. 2000a; Kong et al. 2000).
In the Very High State (VHS), X-ray spectrum is a combination of the disk black-body 
($kT \sim 1.2$keV)  and photon spectral index is $\alpha \sim 2.5$. 
In the High State (HS), X-ray spectrum is dominated by the disk blackbody 
component ($kT \sim 1$keV) and the power law component ($\alpha \sim 2-3$ is weak or absent.
In the Intermediate State (IS) X-ray spectrum shows both power-law ($\alpha \sim 2.5$) and disk blackbody
component ($kT \leq 1$keV). In the Low State (LS) the $2-10$keV photon spectral index 
is around $\alpha \sim 1.5$ and sometimes very weak disc black body component
($kT \leq 1$keV) may be observed. In the Off State (OS) the X-ray fast time variability 
is similar to that of an LS, but the $2-10$keV flux is about 10 times lower or even fainter 
compared to LS. In the case of objects like GRS 1915+105, a few other states are 
in the literature: In the Burst-Off State (BOffS) the spectrum is 
hard, but not as hard as in a LS or OS. Here the count rate is low and 
low Hardness ratio HR1 (Belloni et al 2000a), and variable HR2 
depending on the length of the event. In the Burst-On State (BOnS), 
count rate is high and high HR1 is seen. A transition from BOffS to BOnS
always accompany a lowering of the count with low HR1 and HR2 (Belloni et al.
2000) and this is called a Dip State (DS).

It is to be noted that a very important and tight correlation between 
the average frequency of the quasi-periodic oscillations (QPO)
and the duration of the so-called `burst-off states' exists, all
aspects of which could be explained satisfactorily when outflows are assumed 
to be present (Chakrabarti \& Manickam 2000; hereafter CM00). They discuss that 
in the `burst-off' states the winds loss out matter from the CENBOL, while in `burst-on
states', the subsonic region (the so-called sonic sphere) is cooled down. 
The driving force is thus lost, and matter falls back on the CENBOL, 
while at the same time the CENBOL size is reduced due to turbulent viscous effects
(see, Chakrabarti and Nandi, 2000 fo rdetails). 
Figure 1(a-b) shows a cartoon diagram of the flows in the `burst-off' (a)
and `burst-on' states (b). A similar correlation was found between the
centroid frequency and the duration by Trudolyobov et al. (1999a) in
some of the days of observation.

\begin {figure}
\vbox{
\vskip 8.0cm
\hskip -8.0cm
\centerline{
\psfig{figure=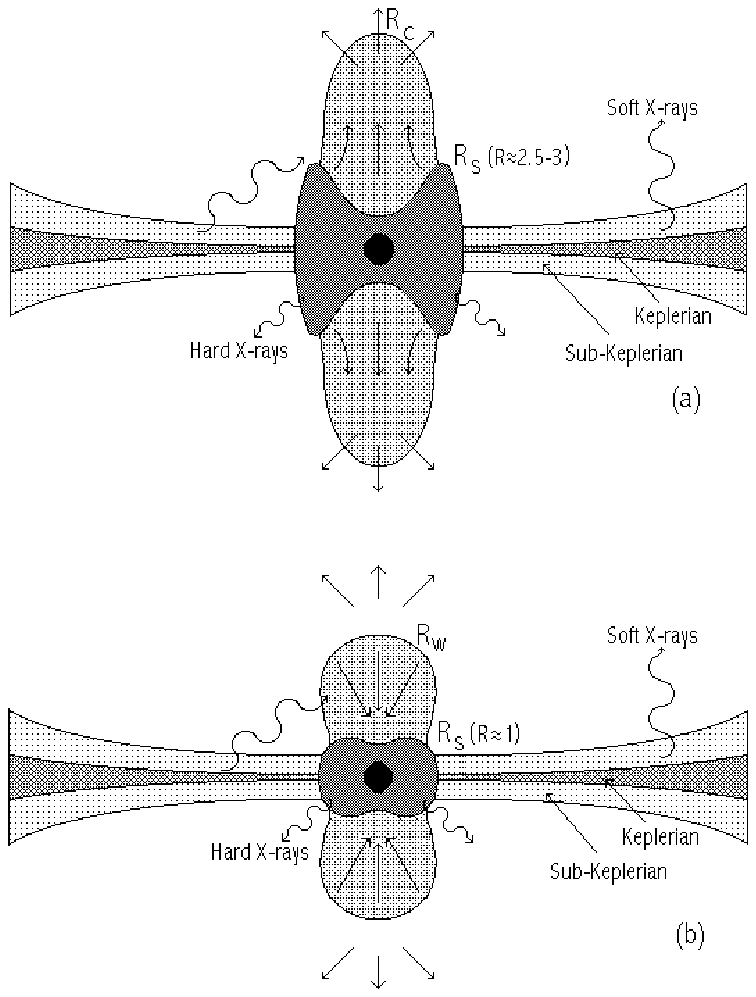,height=10truecm,width=8truecm}}}
\end{figure}
\begin{figure}
\vspace{-9.0cm}
\caption[] {A cartoon of the accretion disk near a black hole
which includes a boundary layer. Fig 1a shows the `burst-off' state where the outflow
material fills up the region upto the sonic point R$_c$. Fig 1b shows the cases with
prominent `burst-on' state where the material in the subsonic region cools and
falls back on the CENBOL (see text) and takes time to drain.}
\end{figure}

What could be the spectral signature of a CENBOL which is losing matter 
to winds?  On the one hand, one might imagine that a CENBOL with lesser 
electron density and therefore optical depth would be hotter and harder. 
But, in presence of a large number of soft photons from the pre-CENBOL flow, 
fewer electrons of CENBOL would be easier to cool. Using this argument, 
Chakrabarti (1998a) computed the spectrum taking loss of matter due to the winds from CENBOL 
into account and showed that emergent spectrum become softer. 
Similarly, when cooler matter of the subsonic region of the outflow falls back (Fig. 1b)
by a return flow, enhanced optical depth of the CENBOL makes it harder to cool it down 
by the same number of soft-photons emerging from the Keplerian 
disk in the pre-CENBOL region. This means that in the burst-on state, 
the spectrum should be hardened in comparison to 
the spectrum in the high/soft state. Thus, {\it softening of the hard state spectrum
and hardening of the soft state spectrum could indicate that 
accretion rate is not preserved within CENBOL-- i.e., matter is 
lost or gained.} (We refer this as `SH-HS' phenomenon for brevity.)
Of course, the degree by which the spectrum changes depends on mass loss/gain.
Since the drainage time of this matter which falls back
is directly related to the duration of the `burst-on'
states (CM00), spectra of both burst-off and burst-on states are affected
and as a result the energy of the pivotal point (where spectra of the hard and soft states intersect)
increases. That the spectral pivoting takes place during normal low to high state excursions
is well known, see, e.g., Ebisawa et al. (1994). What is new in our analysis 
is the observation of the significant shifting of the pivoting point
in cases which exhibits prominent burst-on states.  In the event burst-on state is 
not prominent and noisy, the spectra of the  off states should still 
be softened. In other words, whenever wind is involved, spectra must 
exhibit SH-HS phenomena. Given that the outflow model enumerated above sharply differs from the
earlier conception of jet formation (which stipulates that matter must 
emerge from regions all over the disk, see, Begelman, Blandford 
and Rees, 1984 and references therein) which would have 
no obvious effect on hardening/softening of the spectra and resultant
shifting of the pivotal energy location, it would be 
interesting to see if GRS1915+105 data shows evidence of SH-HS phenomena.
The question becomes very relevant since there are increasing evidence today 
(e.g., Junor, Biretta and Livio 1999) that jets do emerge 
within tens to hundreds of Schwarzschild radii of a black 
hole for the active galaxy M87. Thus if one also observes 
similar behaviour for GRS1915+105, it would justify 
the term micro-quasar from all practical points of view. 

In this {\it Paper}, we present the RXTE spectra of 
GRS1915+105 and show that there is a distinct signature of 
mass loss in the burst-off state and mass gain (fall-back of matter)
in the burst-on state of the spectra. As a result, when both of these
states are prominent, the pivotal energy during burst-off and burst-on 
transitions is shifted towards much higher energy 
than where it would have been had there been {\it direct} low-high transition.  
Even when these burst-on/burst-off states are not prominent SH-HS phenomena 
are still observed. These point to the fact that the hard X-ray 
emitting region, namely, the inner part of the advective 
disks, are directly responsible for producing the winds as well. This 
conclusion is in line that of other observers that the Compotinizing photons
originate from the base of the jet (e.g., Eikenbery et al. 1998; 
Belloni et al. 2000b; Fender and Pooley, 2000).
At this point we must point out that it is not the aim of this paper to
prove that the theoretical prediction on the spectral slope 
is valid for {\it all} the observations made on {\it all} days because of 
obvious constraints. However, we shall choose sufficiently general
observational data set and we would hope that the conclusions drawn would be
valid for observations not covered in this paper. In fact, since the physical 
reasons are generic, we believe that the conclusions would be valid 
even objects other than GRS1915+105. These would be discussed separately.

\section{\bf RXTE PCA Spectral Data }

The RXTE public archive contains several observations on GRS~1915+105 
obtained using the RXTE Proportional Counter Array (PCA - see Jahoda et al. 
1996 for a description of PCA). These observations typically last for a few
thousand seconds and the observations are carried out roughly once a week.
The source was in a low state from December, 1996 up to April, 1997.
The spectral and temporal behavior during the low state was stable 
characterized by a hard spectrum (with the power-law photon index
of $\sim$2, and the total flux in the power-law component being
$\sim$80\%) and $1-10$ Hz QPOs (Trudolyubov et al. 1999b; Muno et al. 1999).
The fact that the canonical low states of black hole
candidate sources have a negligible thermal component
(Chitnis et al. 1998) prompted Trudolyubov  et al. (1999b) to characterize this
state as an ``intermediate state'' and they conclude that with the
lowering of the accretion rate the source should go to the canonical
hard state. Since the source was in similar state on several
occasions (1996 July-August; 1997 October; 1998 September-October),
we treat this state as the low state of GRS~1915+105. The source reached 
a high state in 1997 August. We have selected one observation 
each from the low state and high state to quantify the
spectral parameters. The low state observation was  carried out on 1997 March 26
with PID number of 20402-01-21-00 (spectral and temporal parameters 
for this observation has been reported in Muno et al. 1999 and Trudolyubov et al. 1999b).
The high state observation was carried out on 1997 August 19 with 
a PID number of 20402-01-41-00 (Muno et al. 1999). Spectra of these two 
states are shown in Fig. 2(a-c) and are marked.

\begin {figure}
\vbox{
\vskip -10.0cm
\hskip 0.0cm
\centerline{
\psfig{figure=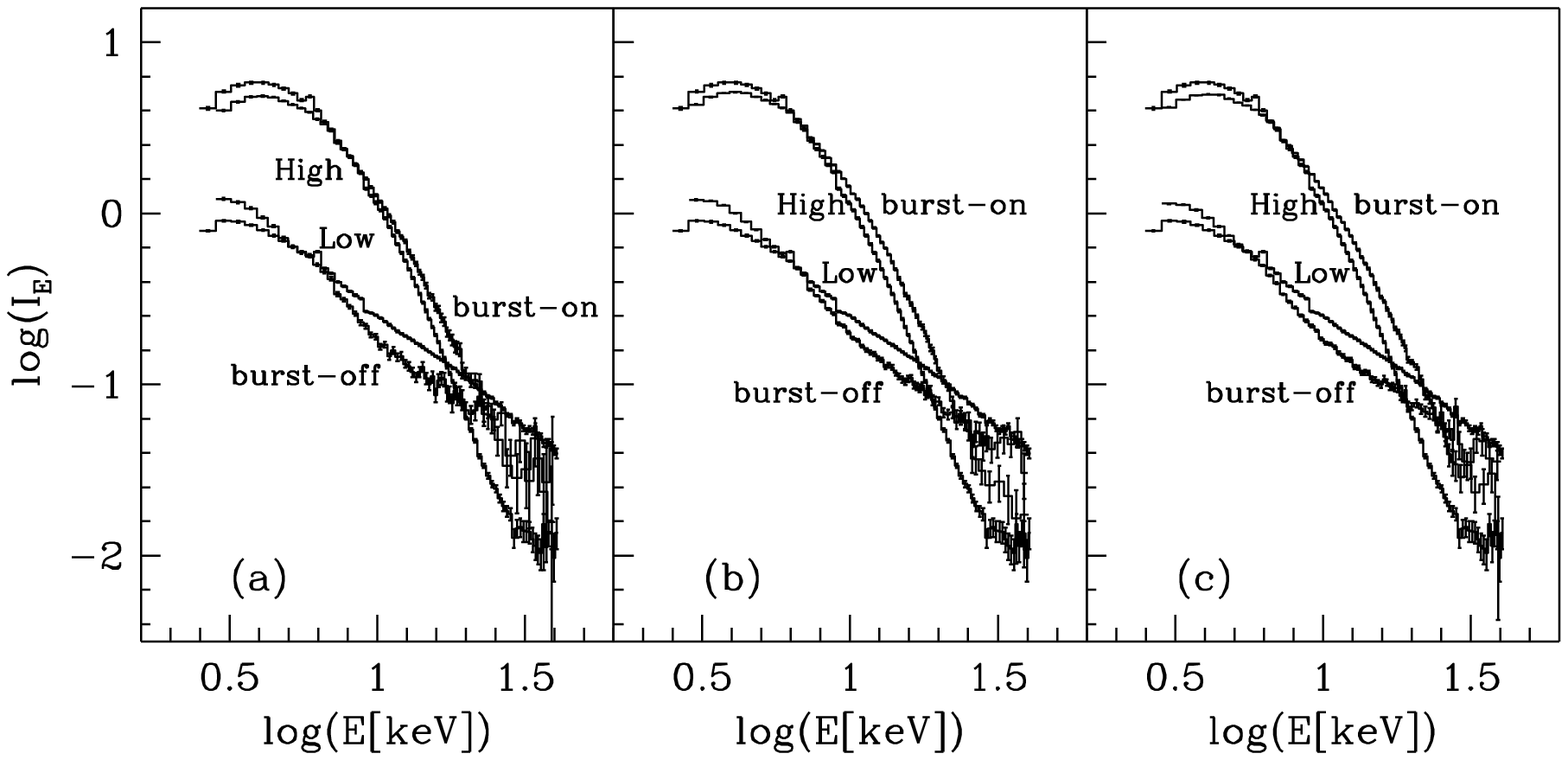,height=18truecm,width=15truecm}}}
\end{figure}
\begin{figure}
\vspace{-1.0cm}
\caption[]
{The unfolded RXTE-PCA spectra of GRS~1915+105 obtained during the low and high states
are compared with burst-on and burst-off spectra during the irregular bursts observed on (a) 1997 June 18,
(b) 1997 July 10, and (c) 1997 July 12. The histogram shows
the fitted model. Due to SH-HS phenomena, during the burst-off/burst-on transition the
pivoting occurs at a higher energy. }
\end{figure}

During the transition between these two states, the source exhibited
several types of bursting behaviour. A detailed discussion of
the classification of light curves are in Yadav et al. (1999), Belloni et al. (2000a)
and Nandi, Manickam \& Chakrabarti (2000).
It was seen that for about a month the source was in a burst mode with a slow transition from
regular bursts to irregular bursts and then again to a regular burst 
of shorter duration (Yadav et al. 1999). In Yadav et al. (1999) 
it was postulated that the irregular bursts are manifestations of 
rapid state changes. As discussed earlier, Chakrabarti and Manickam (2000)
explained the burst-off/burst-on transitions in terms of the repeated 
filling of the outflow region and its abrupt cooling due to inverse Comptonization.
We present the spectral properties of the source during the irregular 
bursts observed on 1997 June 18 (PID 20402-01-33-00) in Fig. 2a,
on 1997 July 10 (PID 20402-01-36-00) in Fig. 2b and on 1997 July 12 
(PID 20402-01-37-01) in Fig. 2c.

We have generated the 129 channel energy spectra from the Standard 2 mode
of the PCA for each of the above observations. Standard procedures for
data selection, background estimation and response matrix generation
have been applied. PCA consists of five units and data from all the 
units are added together. We have fitted the energy spectrum of the 
source using a model consisting of disk-blackbody and power-law with 
absorption by intervening cold material parameterized as equivalent 
Hydrogen column density, N$_H$. The value of N$_H$ 
has been kept fixed at 6 $\times$ 10$^{22}$ cm$^{-2}$.
We have included a Gaussian line near the expected K$_\alpha$  emission
from iron and absorption edge due to iron. These features help to mimic
the reflection spectrum usually found in other Galactic black hole
candidate sources like Cygnus X-1 (Gierlinski et al. 1997).
Systematic errors of 1 $-$ 2 \% have been added to the data.
XSPEC version 10.0 has been used to fit the spectra.

The resultant unfolded spectra for the two spectral states presented in 
Figures 2(a-c) show that in the low state, the disk blackbody component has 
lower temperature (kT$_{in}$ = 0.60$\pm$0.05 keV) and a larger inner disk 
radius (R$_{in}$ = 115$\pm$2 km) compared to the high state, which has 
the inner disk temperature of 1.95$\pm$0.01 keV and the inner disk radius 
of 26$\pm$1 km. The inner disk parameters are calculated using a distance 
to the source of 10 kpc and inclination to the disk of 70$^\circ$. 
The power-law index in the low state (2.40$\pm$0.01) is much flatter 
than that seen in the high state (3.61$\pm$0.02). The quoted errors 
are for nominal 90\% confidence levels obtained by the condition of 
$\chi_{min}^2$ + 2.7. The disk blackbody component has a 3 $-$ 26 keV 
flux of $<$10\% of the total flux in the low state, which increase to $>$65\%.
The two spectra intersect at around 17 keV. It should be noted here
that the disk-blackbody inner radius gives systematically lower value
when scattering effects are not considered and PCA generally shows steeper
spectrum due to uncertainties in the response matrix. The results presented
here, however, highlights the broad changes in the spectral states.

The corresponding unfolded spectra during the irregular burst of June 18, 1997
presented in Figure 2a shows that the burst-off state has spectral parameters
similar to that of the low  state (kT$_{in}$ = 0.76 keV, 
index = 2.76) and the burst-on state spectrum resembles that
of the high state (kT$_{in}$ = 2.2  keV, index = 3.1). Atthe same time
it is clear that the 
spectrum of the burst-off state is softer than that of the low state and the spectrum of the 
burst-on state is harder than that of the high/soft state. As a result, the energy at which 
the two spectra pivot is much high ($\sim$ 26 keV). The same behaviour is
seen also in Fig. 2b and 2c where the IDs used were 20402-01-36-00 and 20402-01-37-01.
The low and high states have been kept as above. The intersection in all these cases
during the burst-on/burst-off transition is around 25 keV ---  
far above the low/high intersection.

In a recent paper, Dhawan et al. (2000) pointed out a direct correlation 
between the ASM X-ray data from RXTE and IR/Radio observations. The 
radio activity at around 500AU made on 31st October, 1997 would  
be perturbed by CENBOL activity of around 28.5 October, 1997, 
if perturbation propagates with $0.98c$. Unfortunately, no PCA 
data is available for 28th or 29th of October. Figure 3 shows  
the spectral properties on 30th October, 1997. This is drawn with 
observation ID 20402-01-52-02. To compare with the high/low spectrum, 
the low state observation closest to this day, namely, of 25th of 
October, 1997, we chose the same high state as mentioned above.
This latter observation ID is 20402-01-52-00. The spectral slopes 
of the high, low, on and off states are 3.61$\pm$0.02, 
2.76$\pm$0.018, 3.34$\pm$0.021 and 2.85$\pm$0.013, respectively.
This again shows the softening of the hard and hardening of the 
soft states. The intersection of the spectra of low and high 
states is at around $14$keV whereas that of  the on and off 
states is at around $17$keV. Thus the effect of winds is still 
prominent even after one and a half day of ejection of matter, 
though the effect is lesser.
 
\begin {figure}
\vbox{
\vskip -5.0cm
\hskip 0.0cm
\centerline{
\psfig{figure=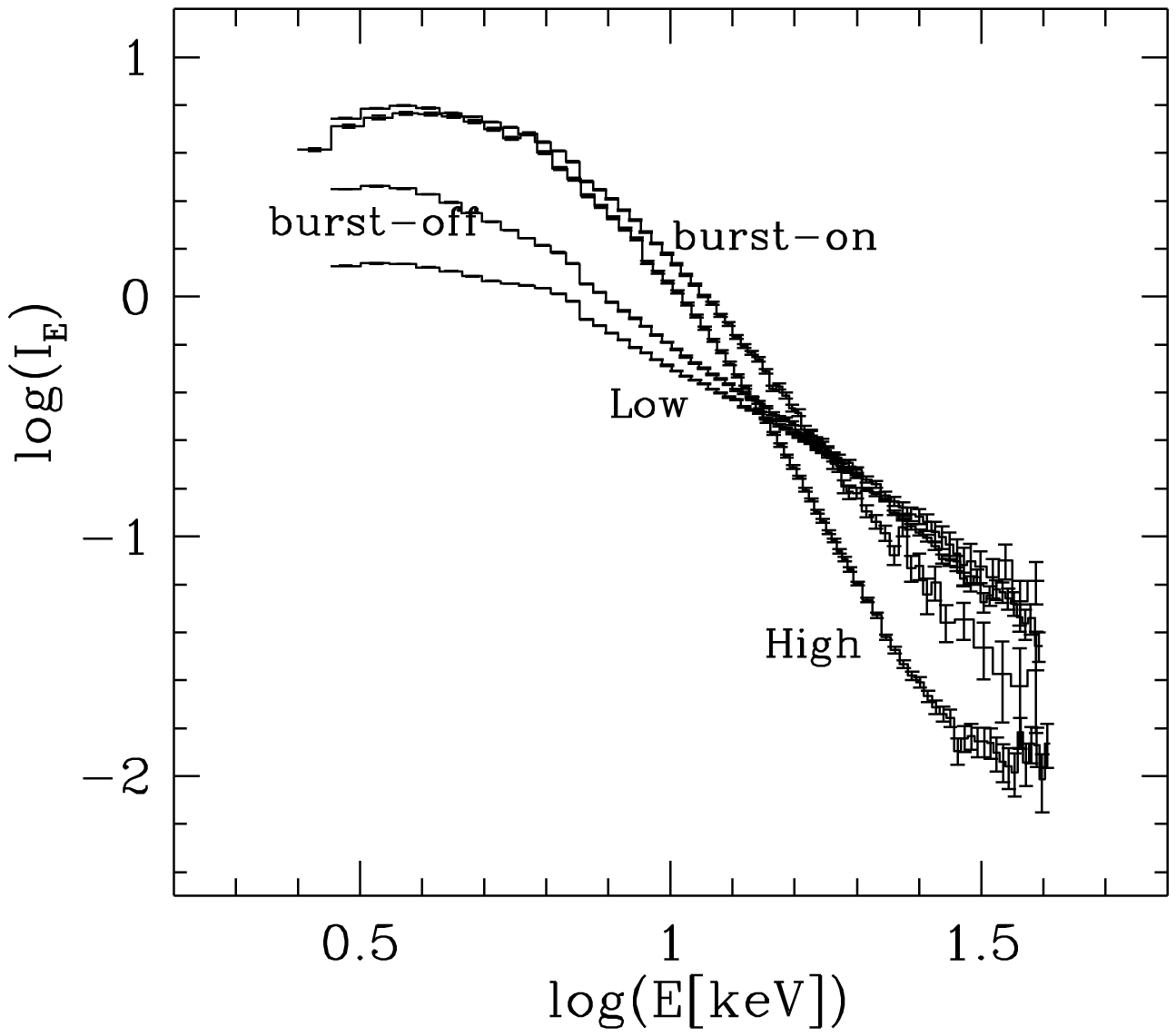,height=15truecm,width=10truecm}}}
\end{figure}
\begin{figure}
\vspace{-3.0cm}
\caption[]
{As in Fig. 2 except that the observation ID of the burst-on/burst-off and low states
are chosen close to 31st October, 1997 when IR/Radio flares were observed (Dhawan et al., 2000).
SH-HS phenomena are seen on this day also. See text for details.}
\end{figure}

\section{Theoretical Interpretations}

As already discussed in the Introduction, the SH-HS phenomena, i.e., 
hardening of the soft-state spectrum and softening of the hard-state spectrum 
and the resulting shift towards higher energy in the pivotal point could be
understood in terms of the  presence of winds (Chakrabarti 1998a) or 
return flow of additional cooler matter to  CENBOL (Chakrabarti and Nandi, 2000) (Fig. 1ab).  
In Fig. 4, we present two sets of calculated spectra to illustrate this. Calculations are 
done using models presented in CT95 and it subsequent modification by  Chakrabarti 
(1997). Disk accretion rates for the soft and the hard states are $0.3 {\dot M}_{Edd}$ 
and $0.1 {\dot M}_{Edd}$ respectively. Halo accretion rate is kept fixed at
$1.0 {\dot M}_{Edd}$ and the shock is located at $X_{s}=14$ and $X_s=10$
respectively in low and high states. This is in line with the general 
conclusions that the inner edge moves in during soft states. Other 
parameters are kept identical. The solid curves are drawn to mimic the 
burst-off and burst-on state spectra. In these cases the disk accretion rates are kept
as before, but the twenty percent of CENBOL matter is assumed to be lost
in wind in the burst-off state and ten percent of matter is assumed to be
falling back on the halo from the wind region in the burst-on state.
Because of selective softening and hardening, the intersection point 
is located at a much higher energy exactly as observed.
After inclusion of these models in XSPEC, in future, it would be possible to
fit the spectra to obtain the flow and the wind parameters.

\begin {figure}
\vbox{
\vskip -5.0cm
\hskip 0.0cm
\centerline{
\psfig{figure=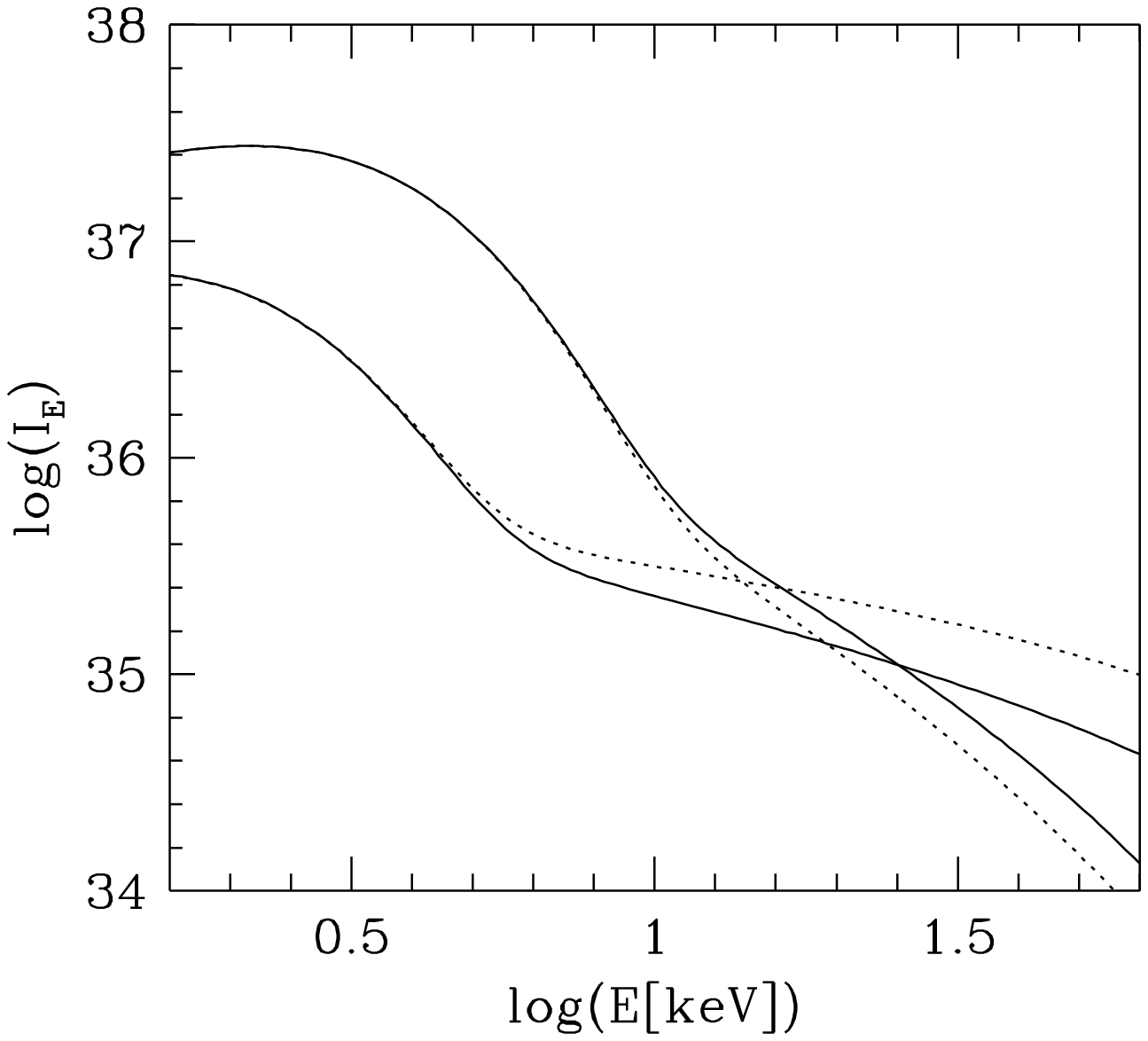,height=15truecm,width=10truecm}}}
\end{figure}
\begin{figure}
\vspace{-3.0cm}
\caption[]
{Typical nature of calculated spectra (uncorrected for absorption)
for burst-off and burst-on states and low and high states of GRS1915+105.
In burst-off state, spectrum appears to be softened with respect to low state, and in burst-on state,
spectrum appear to be hardened with respect to the high state.
See text for details.}
\end{figure}

\section{Discussion and conclusions}

A comparison of the two spectral states in Fig. 2(a-c) and Fig. 3 shows that
during the  bursting phase the low energy component ($\lsim$ 5 keV) remains essentially
same as that found during the extended low and high states of the
source. The major change is above $\sim$ 10 keV where the hard component
of the low  state softens  to  form a burst-off state and that of the high state hardens to form 
a burst-on state. Particularly interesting is the result in Fig. 3, where
actual observations of outflow is present (Dhawan et al. 2000). 
We presented a possible scenario where we incorporated recent findings that loss of matter
in winds might cause the spectral softening. We showed earlier that 
winds are lost from the centrifugal barrier in the burst-off state of the flaring
phase and a part of the matter falls back to increase the sub-Keplerian flow  
thereby hardening the spectra during the burst-on state. It has already been
argued (Chakrabarti, 1999a) that the wind formation rate may be dependent
on the compression ratio of the flow at the shock and that the outflow
rates are related to the spectral states (Chakrabarti, 1999b). These results are
fully in line with multi-wavelength observations of correlated variabilities
(Mirabel \& Rodriguez 1999; Dhawan et al, 2000; Fender and Pooley, 2000; 
Belloni, Migliari and Fender, 2000). Our present result where  matter 
loss/gain is seen to affect the {\it Comptonized} photons,
thus indicates that winds are produced from the `boundary layer' of a black hole.

The connection between winds from the accretion disk, the AU scale jets
in the core of GRS~1915+105 (Dhawan et al. 2000) and the super-luminally
moving synchrotron ejecta (Mirabel \& Rodriguez 1994; Mirabel \& Rodriguez 1999;
Fender et al. 1999b) are not very clear at present and a detailed evaluation is beyond the
scope of the present paper. The accretion-wind scenario presented in
Chakrabarti (1999a) and Das \& Chakrabarti (1999) gives a self-consistent
way to generate winds from accretion disks, but the acceleration of
such winds into superluminally moving ejecta is a separate matter.
We can conjecture that GRS 1915+105 has a CENBOL location such
that most of the times there is a strong wind emission. This wind
can be confined in the sonic sphere (to give rise to the
repeated outbursts), can be accelerated steadily (to give flat
spectrum radio emission) or ejected at superluminal velocities
(to give rise to the steep spectrum synchrotron ejecta) depending on initial 
parameters (which in turn govern the shock compression ratio). 
The basic mechanism of the wind emission from the CENBOL 
and the consequential change in spectral slopes
are not inconsistent with any known observations. A crucial aspect of this
model is the scattering of the soft photons by modified opacity of the CENBOL: 
a subtle difference that has been confirmed in the present work. 
Due to obvious constarint we could notcheck our conjecture for all the days 
of observations or for other objects, but we believe that it would remain valid. 
The exact mechanism of the acceleration (steady as well as superluminal
motion) probably requires some magneto-hydrodynamic mechanism.

\begin{acknowledgements}

This research has made use of data obtained through the High Energy
Astrophysics Science Archive Research Center Online Service, provided by the
NASA/Goddard Space Flight Center.  SKC and SGM acknowledge the support of 
Indian Space Research Organization for a RESPOND project. AN acknowledges 
DST project grant No. SP/S2/K-14/98 for funding his research at SNBNCBS.
\end{acknowledgements}

{}
\end{document}